\documentclass[conference]{IEEEtran}

\usepackage[cmex10]{amsmath}

\usepackage{epsfig,amssymb,amsfonts,url}

\def\mylabel#1{\label{#1}}

\allowdisplaybreaks[1]

\begin{document}

\title{Loss Fluctuations and Temporal Correlations in Network Queues}

\author{\IEEEauthorblockN{I.~V.~Lerner and I.~V.~Yurkevich}
\IEEEauthorblockA{School of Physics and Astronomy \\ University of Birmingham, Edgbaston \\ Birmingham, B15 2TT, UK\\
Email: \{i.v.lerner,i.yurkevich\}@bham.ac.uk}
\and
\IEEEauthorblockN{A.~S.~Stepanenko and C.~C.~Constantinou}
\IEEEauthorblockA{School of Engineering \\ University of Birmingham, Edgbaston \\ Birmingham, B15 2TT, UK\\
Email: \{a.stepanenko,c.constantinou\}@bham.ac.uk}}

\IEEEspecialpapernotice{(Invited Paper)}

\maketitle

\begin{abstract}
We consider data losses in a single node  of a packet-switched Internet-like network. We employ two distinct models, one with discrete and the other with
continuous one-dimensional  random walks, representing the state of a queue in a router. Both models
{have} a built-in critical behavior with {a sharp} transition from exponentially small to finite losses. It turns out that the finite capacity of a buffer and the packet-dropping procedure give rise to specific boundary conditions which lead to strong
  loss rate
fluctuations at the critical point even in the absence of such fluctuations in the data arrival
process.
\end{abstract}

\section{Introduction}

Many systems, both natural and man-made are organized as complex networks of  interconnected entities: brain cells
\cite{Arbib:01}, interacting molecules in living cells \cite{Jeong:00}, multi-species food webs \cite{Cohen:90}, social
networks \cite{Liljeros:01} and the Internet \cite{Pastor:01} are just a few examples. In addition  to the classical
Erd\"{o}s--R\'{e}nyi model for random networks  \cite{ErdRen}, new overarching models of scale-free \cite{Barabasi:99}
or small-world \cite{Watts:98} networks turn out to describe real world examples. These and other network models have
received extensive attention by physicists (see Refs.~\cite{Watts:99,Albert:02} for reviews).

A particularly interesting problem for a wide range of complex networks is their resiliency to breakdowns. The
possibility of random or intentional breakdowns of the entire network has been considered in the context of scale-free
networks where nodes were randomly or selectively removed \cite{Albert:00,Cohen:00,Braunstein:03}, or in the context of
small-world networks where  a random reduction in the sites' connectivity leads to a sharp increase in the optimal
distance across network which destroys its small-world nature \cite{Braunstein:03,Dorogovtsev:00,Ashton:05}. In all
these models, the site or bond disorder acts as an input which makes them very general and applicable to a wide variety
of networks.

Network breakdowns    can  result not only from a physical loss of connectivity but  from an operational failure of
some network nodes to forward data. In the more specific class of communication networks, this could happen due to
excessive loading of a single node. This could trigger cascades of failures and thus isolate large parts of the network
\cite{Moreno:03}. In describing the operational failure in a particular network node, one needs to account for distinct
features of the dynamically `random' data traffic which can be a reason for such a breakdown.

In this paper we model data losses in a \textit{single node} of a packet-switched network like the Internet. There are
two distinct features which must be preserved in this case: the discrete character of data propagation and the
possibility of data overflow in a single node.  In {the packet-switched network} data is divided into packets which are
routed from source to destination via a set of interconnected nodes (routers). At each node packets are queued in a
memory buffer before being {\it serviced}, i.e.\ forwarded to the next node. (There are separate buffers for incoming
and outgoing packets but we neglect this for the sake of simplicity). Due to the finite capacity of memory buffers and
the stochastic nature of data traffic, any buffer can become overflown which results in  packets {being {\it
discarded}}.

{We consider a model} where {noticeable} data losses in a single memory buffer start when the average rate of random
packet arrivals approaches the service rate. {Under this condition} the model has {a built-in sharp} transition from
free flow to lossy behavior  {with} a finite fraction of arriving packets {being} dropped. A sharp onset of network
congestion is familiar to everyone using the Internet and was numerically confirmed in different models
\cite{Ohira:98}. Here we stress that such a congestion can originate from a single node.

{While data loss is natural and inevitable due to the data overflow, we show that loss rate statistics  turn out to be
highly nontrivial in the realistic case of a finite buffer}, where at the critical point  the magnitude of fluctuations
can exceed the average value, while they obey the central limit theorem only
 in the (unrealistically) long time limit. Such an
importance of fluctuations in some intermediate regime is a definitive feature of {\it mesoscopic} physics, albeit the
reasons for this are absolutely different (note that even in the case of electrons, the origin of the mesoscopic
phenomena can be either quantum or purely classical, see, e.g., \cite{IVL:93}). Although we model data arrivals as a
Markovian process, the loss rate at intermediate times shows long-range power-law correlations in time. When excessive
data losses start, it is more probable that they persist for a while, thus impacting on network operation.

The {\it average} loss rate and/or transport delays were previously studied, e.g.,  in the {theories} of bulk queues
\cite{Cohen:69,Schwartz:87} {or a jamming transition in traffic flow} \cite{Cates:98}. What makes present
considerations qualitatively different is that we analyze {\it fluctuations} of a {\it discrete} quantity, the number
of discarded packets. Although fluctuations in network dynamics were previously studied  (see, e.g.\ \cite{Menezes}),
this was done in {the} continuous limit for the data traffic,  through measurements or numerical simulations.

\section{The Discrete Model}

The mode of operation of a memory buffer is that packets arrive
randomly, form a queue in the buffer  and are subsequently {\it
serviced}.  Each packet in the queue has typically a variable
length and is normally serviced in fixed-length service units  at
discrete time intervals on a first-in first-out basis. Here we
choose the simplest non-trivial model of this class:  (i) packets
have a fixed length of two service units;  {(ii) arrival and
service intervals coincide}. The length of the queue after $n$
service intervals, $\ell_n$, serves as a dynamical variable {which
obeys the discrete-time Langevin equation,}
\begin{align}\label{L}
    \ell_{n+1}=\ell_{n}+\xi_{n}\,,
\end{align}
{where the telegraph noise $\xi_{n}$ is defined by}
\begin{align}\notag
    \xi_n&= \left\{%
\begin{array}{lc}
    1, &{0\le\ell_n\le L-1} \\
    0, &{\ell_n=L}\,, \\
\end{array}%
\right. {\text{ with  probability $p$}}
 \\[-4pt] \label{1}\\[-4pt]\notag
    \xi_n&= \left\{%
\begin{array}{lc}
       \phantom-0, &{\ell_n=0}\,, \\
 -1, &{1\le\ell_n\le L} ,
\end{array}%
\right.{\text{ with  probability $1-p$}}
\end{align}
{The above means that} the length of the queue, {measured in
service units}, either increases by one   when one packet arrives
and one service unit is served, or decreases by one when no packet
arrives. The  boundary conditions above correspond to discarding a
newly arrived packet when buffer is full ($\ell_n=L$) and to an
idle interval when no packet arrives at an empty buffer
($\ell_n=0$). In a more general, continuous model we will remove these restrictions, allowing for  arbitrary quasi-Markovian nature of the input data traffic. The characteristic features of our results would not change.

\begin{figure}[t]

\begin{center}
\leavevmode \epsfxsize=0.4\textwidth \epsffile{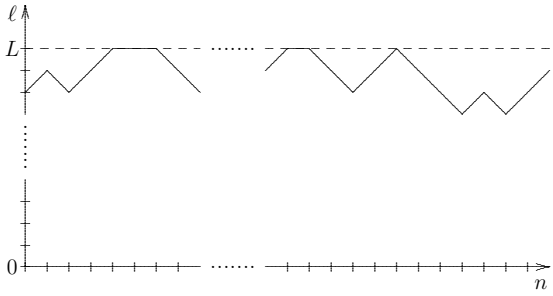}
\end{center}

\caption{The model of data losses: incoming packets randomly
arrive in discrete time intervals and join the queue of length
$\ell$ limited by the memory buffer capacity $L$. Packets in front
of the queue are served at the same time intervals. If the queue
sticks to the  boundary, newly arriving packets are discarded.}

\label{Figure 1}

\end{figure}

The main quantity which characterizes congestion is the packet loss rate
which is defined via the number of packets discarded during a time
interval $N$ by
\begin{align} \label{loss}
\mathcal{L}_N(n_0)= \sum_{n=n_0+1}^{n_0+N} \delta_{\ell_n,
L}\delta_{\ell_{n+1}, L}\,.
\end{align}
This means   that the  packet is discarded if by the moment of its
arrival the queue was at the maximal capacity $L$ as illustrated
in Fig.~\ref{Figure 1}. Thus the loss rate (\ref{loss}) is defined
entirely by the processes  at the boundary of the random walk (RW)
so that the continuous limit cannot be exploited. This makes the
loss statistics profoundly different from, e.g., the thoroughly
studied statistics of first-passage time.

We will consider the  average and the variance of the loss  rate, which we obtain directly from Eq.~(\ref{loss}):
\begin{align}
     \label{l-av}
   & \left<{{\mathcal{L}}_N}\right>=
 \mathcal{P}_{\text{st}}(L)N{\mathcal{G}_1(L,L)}\,,
\\
     \label{l-var}
    &\left<{{\mathcal{L}_N}^2}\right>= \sum_{n,m=n_0+1}^{n_0+N}\left<{
    \delta_{\ell _n,L} \delta_{\ell _{n+1},L}
   \delta_{\ell _m,L} \delta_{\ell _{m+1},L}
    }\right>\notag\\
    &=\left<{{\mathcal{L}}_N}\right>+2{\mathcal{P}_{\text{st}}}(L){\mathcal{G}}_1^2
    (L,L)\sum_{n<m}^{}{\mathcal{G}}_{m-n-1}(L,L)\,.
\end{align}
where the conditional probability of the queue being of
length $\ell$ at time $n$ provided  that it was of length $\ell'$
at time $ n_{0}$ is defined by
\begin{align*}
{\mathcal{G}}_{n-n_0}({\ell},{\ell}')=\left<{\delta_{{\ell}_n
,{\ell}}\, \delta_{\ell_{n_0},{\ell}'}}\right>/\left<{
\delta_{\ell_{n_0},{\ell}'}}\right>\,.
\end{align*}
Here $\left<{\ldots}\right>$ stand for the averaging over the
telegraph noise of Eqs.~(\ref{L}) -- (\ref{1}).  The stationary
distribution of the queue length is related to ${\mathcal{G}}$ by
\begin{align} \label{st}
    \mathcal{P}_{\text{st}}(\ell)=\lim_{n_0\to-\infty}
    {\mathcal{G}}_{n-n_0}(\ell,\ell')=
\left<{\delta_{{\ell}_n ,{\ell}}} \right>\,.
\end{align}
After relatively straightforward calculations \cite{Yur06}, we find and expected (actually, built-in) critical behavior of the average:
\begin{align}
     \frac{1}{N}\left<{{\mathcal{L}}_N}\right>&=
    p\frac{q^{L+1}-q^L}{q^{L+1}-1} \;  \begin{array}{c}
        \\[-8pt]
      \longrightarrow \\[-8pt]
      _{L\gg1} \\
    \end{array}\;
    \left\{%
\begin{array}{ll}
    2p-1, & {p>\frac12;} \\[6pt]
    \frac{1}{L+1}, & {p=\frac12;} \\[6pt]
    \frac{1-2p}{1-p}\,q^L, & {p<\frac12;} \\
\end{array}%
\right.\label{AV}
\end{align}
where $q\equiv p/(1-p).$
Thus the loss rate  for $p>1/2$ is of order 1, for $p=1/2$ a small
fraction of the buffer capacity and for $p<1/2$ an exponentially
vanishing function, as expected. The matching between the three
asymptotic regimes takes place in a narrow region (of width
$\sim1/L$) around $p=\frac{1}{2}$.

\begin{figure}[t]

\begin{center}
\leavevmode \epsfxsize=0.45\textwidth \epsffile{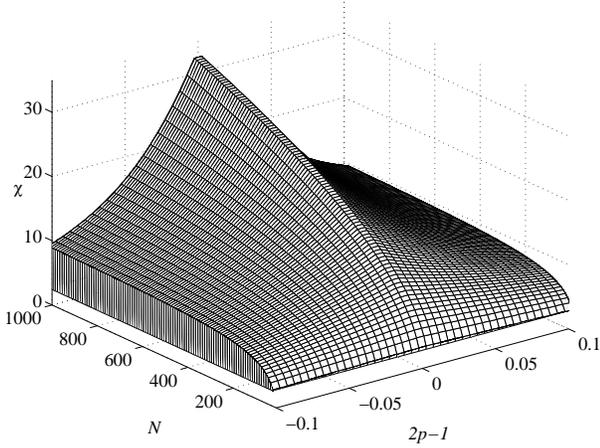}
\end{center}

\caption{compressibility $\chi$ (for $L=1000$) shows a fast
increase of fluctuations in time  at the critical point, $p=1/2$.
}

\label{Figure 2}

\end{figure}

The result for the variance is far more interesting, as it shows significant fluctuations in the loss rate.
It
  is convenient to
express the variance in terms of the `compressibility' defined by
\begin{align} \label{var-chi}
     \langle \delta{\cal L}^2_N\rangle&\equiv\chi_N\,\langle
{\cal L}_N\rangle\,, &\delta\mathcal{L}_N(n) &= \mathcal{L}_N(n)-
\left\langle
     \mathcal{L}_N\right\rangle\,.
\end{align}
The  compressibility has been exactly calculated in \cite{Yur06}. Its behavior of $\chi$ is illustrated in Fig.~\ref{Figure 2} which shows
its fast increase at the critical point, $p=1/2$.

   For $N\gg N_0
   \equiv \left[(2{p}-1)^2+(\pi/L)^2\right]^{-1}\,,
$
i.e.\ in the   limiting
steady-state regime the compressibility saturates at
\begin{align}%
\label{lim}
    \chi_\infty=\left\{%
\begin{array}{ll}
    \frac{1-|2p-1|}{|2p-1|}, &  |2p-1|L\gg1 \\[8pt]
    \frac23L,&|2p-1|L\ll1 \\
\end{array}%
\right.
\end{align}
Although it diverges  in the thermodynamic limit, $L\to\infty$ and
$N/L^2\to \infty$, at the transition point $p=1/2 $, the variance
(\ref{var-chi}) in this limit remains finite and obeys the central
limit theorem.

However, this limit is reachable at the critical point only for
unrealistically long times $N\!\gg \!N_0\!\propto\! L^2$. In the
intermediate, practical regime, $1\ll N\ll N_0$, the
compressibility rapidly increases with time:
\begin{align} \label{chiN}
\chi_{N}&=c\,N^{1/2}, & c&=\frac{2\sqrt
2}{\pi}\int\limits_{0}^{\infty}\frac{dx}
{x^2}\left(1-\frac{1-e^{-x^2}}{x^2}\right),
\end{align}
so that the variance exceeds the average value of the loss-rate
and, as can be shown, its distribution is no longer normal.

It is even more interesting that in
this regime the fluctuations of the loss rate are no longer
Markovian as they exhibit long-time correlations. To show this,
we consider the temporal correlation function of the loss rate
defined by
\begin{equation*}
     R_2(N,M)\equiv
     \frac{\left\langle \delta\mathcal{L}_N(0)\,
     \delta\mathcal{L}_N(M)\right\rangle}%
     {\left\langle \delta\mathcal{L}_N^2\right\rangle}\, ,\ \ M>N\,.
\end{equation*}
 In the most relevant
regime, $N_0\gg N\gg1$ and $M>N$, this has been calculated to give
\begin{align}\notag
     R_2(N,M) = &\frac{p N}{\chi_N}
     \Biggl[ {\rm e}^{-M(2p-1)^2/2}\sqrt{\frac{2}{\pi M}}\\
     &
     - |2p-1|\mathop{\rm erfc}\left(|2p-1|\sqrt{M/2}\,\right)\Biggr].
\end{align}
At the critical point this reduces using Eq.~(\ref{chiN}) to
\begin{equation}\label{R2}
     \left.R_2(N,M)\right|_{p=1/2} = c^{-1}\sqrt{\frac{N}{2\pi M}}\,.
\end{equation}
This long-time correlation (in spite of the packet arrival being
Markovian) is another clear sign of criticality.

\section{The Continuous Model}

Consider the output buffer attached to the switching device in the router. The speed of the input line of the buffer
(effectively, it is the capacity of the switching fabric) would normally be much bigger than the speed of the output
line of the buffer. Hence, we can consider the packet arrival as an instantaneous process. Packets arrivals are treated
as a renewal process. The capacity of the buffer is $L$ (measured in bits). The lengths of arriving packets are treated
as random (measured in bits). The service (the capacity of the output link) is considered to be deterministic. Note
that the packets are put in frames before being sent to the output link, the corresponding randomness (due to the
randomness of packet size) is neglected because it is much smaller of the randomness of the arrival process (the
overheads are much smaller than the typical size of the packet). We normalise the lengths of packets $p$, the speed of
the output link $r_{\rm out}$ and the queue length $\ell$ by the size of the buffer $L$ (the size of the buffer is then
set to be $1$).

The procedure is the following: assume that at the moment of arrival of a packet of size $p$, the state of the queue is
$\ell$, this is followed by the gap $\eta$ (random inter-arrival time) till the next arrival. If $\ell+p\le 1$ then the
packet joins the queue and the queue length prior the next arrival is $\ell'=\ell+p-\eta/\eta_0$ if $\ell'>0$ and
$\ell'=0$ otherwise. If $\ell+p > 1$ then the packet is discarded and the queue length prior the next arrival is
$\ell'=\ell-\eta/\eta_0$ if $\ell'>0$ and $\ell'=0$ otherwise. Here we introduced the following time scale
\begin{equation}
  \eta_0\equiv\frac{1}{r_{\rm out}}\ ,
\mylabel{eta0}
\end{equation}
the time required to empty a full buffer provided there are no new arrivals.

Assuming that the maximum packet size is much less than $1$ (the buffer size) and the average incoming traffic rate
$r_{\rm in}$ (also normalised to the buffer size) is close to the service rate:
\begin{equation}
  | r_{\rm in}\eta_0 - 1 | \ll 1
\mylabel{eq0}
\end{equation}
we can treat $p$, $\eta$ and $\ell$ as continuous variables.

Our aim is to calculate the statistics of the amount of the dropped traffic and the service lost due to idleness of the
output link during time $t\gg\bar\eta$ ($\bar\eta$ is the average inter-arrival time) in the regime (\ref{eq0}).

In the regime (\ref{eq0}) and for observation times $t\gg\bar\eta$, the system can be described by the Fokker-Planck
equation as follows (in terms of the transitional probability density function $w(\ell',t;\ell)$)
\begin{equation}
 \partial_t w(\ell',t;\ell)
  = - a \partial_{\ell'} w(\ell',t;\ell)
    + \frac{1}{2}\sigma^2\partial_{\ell'}^2 w(\ell',t;\ell)\ ,
\mylabel{FP}
\end{equation}
where $a$ and $\sigma^2$ are average moments of the change of the queue size per unit time
\begin{equation}
 a \equiv\frac{1}{\Delta t}\langle \Delta \ell \rangle\ , \quad
 \sigma^2 \equiv\frac{1}{\Delta t}\langle \Delta \ell^2 \rangle\ ,\quad \Delta t\to 0
\mylabel{moments}
\end{equation}
and the following boundary and initial conditions are imposed
\begin{equation}
 \left. J(\ell',t;\ell)\right|_{\ell'=0,1} = 0\ , \quad
\mylabel{boundary}
\end{equation}
\begin{equation}
 \left.w(\ell',t;\ell)\right|_{t=0} = \delta(\ell'-\ell)
\mylabel{initial}
\end{equation}
where
\begin{equation}
 J(\ell',t;\ell)\equiv a w(\ell',t;\ell) - \frac{1}{2}\sigma^2\partial_{\ell'} w(\ell',t;\ell)\ ,
\mylabel{current}
\end{equation}
is the probability current. By $\Delta t\to 0$ in eq.~(\ref{moments}) we mean that it is much smaller than the
observation time, but large enough so that the underlying stochastic processes can be considered as continuous:
\begin{equation}
 \bar\eta\ll\Delta t\ll t
\mylabel{time}
\end{equation}

The solution of (\ref{FP},\ref{boundary},\ref{initial}) can be expressed as follows
\begin{equation}
\begin{split}
 w(\ell',t;\ell)
  =& 2{\rm e}^{v(\ell'-\ell)}\sum\limits_{k=1}^\infty
  \frac{\exp\left[-(4\pi^2k^2 + v^2)\tau\right]}{4\pi^2k^2 + v^2}
\\
&\quad\times
  \left[ 2\pi k\cos(2\pi k\ell') + v \sin(2\pi k\ell') \right]
\\
&\quad\times
  \left[ 2\pi k\cos(2\pi k\ell) + v \sin(2\pi k\ell) \right]
\mylabel{eq1a}
\end{split}
\end{equation}
where
\begin{equation}
 v\equiv\frac{a}{\sigma^2}\ ,\qquad
 \tau\equiv\frac{\sigma^2t}{2}
\mylabel{eq1b}
\end{equation}
Note that the solution (\ref{eq1a}) can be expressed in terms of $\theta$-functions.

For the Laplace transform of $w(\ell',t;\ell)$ we have
\begin{equation}
\begin{split}
 &W(\ell',\epsilon;\ell)
  \equiv {\cal L}_\tau w(\ell',t;\ell)
 = \frac{1}{2}\frac{{\rm e}^{v(\ell'-\ell)}}{\kappa\sinh(\kappa)}
\\
  &\quad \times
  \Biggl\{
     \frac{2v^2}{\epsilon}\cosh[\kappa(\ell'+\ell-1)]
   + \frac{2\kappa v}{\epsilon}\sinh[\kappa(\ell'+\ell-1)]
\\
&\qquad
   + \cosh[\kappa(|\ell'-\ell|-1)] + \cosh[\kappa(\ell'+\ell-1)]
  \Biggr\}
\mylabel{eq2}
\end{split}
\end{equation}
where
\begin{equation}
 \kappa\equiv\sqrt{\epsilon + v^2}
\mylabel{eq3}
\end{equation}
From (\ref{eq2}) we have for the probabilities of returning to the boundaries
\begin{equation}
\begin{split}
 W(0,\epsilon;0) &= \frac{1}{\epsilon}\left[ \kappa\,{\rm cotanh}(\kappa) - v \right]
\\
 W(1,\epsilon;1) &= \frac{1}{\epsilon}\left[ \kappa\,{\rm cotanh}(\kappa) + v \right]
\end{split}
\mylabel{eq4}
\end{equation}
These will be used in the next section.

\section{Statistics of losses}

In this section we will concentrate on the statistics of the losses due to the buffer overflowing. The corresponding
formulae for the statistics of the server idleness can be obtained using transformation $\ell\to1-\ell,v\to-v$.

First, we estimate the size of fluctuations of the losses on time scale $t\ll 2/\sigma^2$. In order to do that we
consider the dynamics of the system near the boundary $\ell=1$ which is governed by the following transitional
probability:
\begin{equation}
\begin{split}
 & w_0(\ell',t;\ell)
 = \frac{1}{\sqrt{2\pi\sigma^2 t}}\exp\left[ - \frac{a(\ell'-\ell)}{\sigma^2} - \frac{a^2t}{2\sigma^2} \right]
\\
&\quad \times
    \left\{
       \exp\left[ - \frac{(\ell'-\ell)^2}{2\sigma^2 t} \right]
     + \exp\left[ - \frac{(2-\ell'-\ell)^2}{2\sigma^2 t} \right]
    \right\}
\\
&\quad
 - \frac{a}{\sigma^2}\exp\left[ \frac{2a(1-\ell')}{\sigma^2} \right]
    {\rm erfc}\left[ \frac{2-\ell'-\ell+at}{\sqrt{2\sigma^2t}} \right]
\end{split}
\mylabel{eq7a}
\end{equation}
which is the solution of (\ref{FP}) when the boundary $\ell=0$ is sent to $-\infty$. The change in the state of the
system during time $t$ can then be represented as follows:
\begin{equation}
 \Delta\ell(t)\equiv \ell' - \ell = \Delta\ell_0(t) + \Delta\ell_{\rm loss}(\ell',t;\ell)
\mylabel{eq6}
\end{equation}
where $\Delta\ell_0(t)$ is the change in the state of the system if there was no boundary, its statistics is determined
by
\begin{equation}
 \langle\Delta\ell_0(t)\rangle = at\ ,\quad
 \langle[\Delta\ell_0(t)]^2\rangle = \sigma^2t + {\rm o}(t)\ ,\quad
\mylabel{eq7}
\end{equation}
and $\Delta\ell_{\rm loss}(\ell',t;\ell)$ is the amount of traffic lost due to buffer overflowing. The moments of
(\ref{eq7}) can be defined as follows
\begin{equation}
 \langle[\Delta\ell(t)]^n\rangle
  = \int\!{\rm d}\ell'{\rm d}\ell\ (\ell'-\ell)^nw_0(\ell',t;\ell)p(\ell)
\mylabel{eq7b}
\end{equation}
where $p(\ell)$ is the stationary distribution.

For the first two moments (\ref{eq7b}) in the limit $t\to0$ we have
\begin{equation}
 \langle\Delta\ell(t)\rangle = at + \frac{\sigma^2t}{2}p(1)\ ,\quad
 \langle[\Delta\ell(t)]^2\rangle = \sigma^2t
\mylabel{eq7c}
\end{equation}
From (\ref{eq6},\ref{eq7},\ref{eq7c}) we can conclude that
\begin{equation}
\begin{split}
 &\langle\Delta\ell_{\rm loss}(t)\rangle = \frac{\sigma^2t}{2}p(1)
\\
 &\langle[\Delta\ell_{\rm loss}(t)]^2\rangle + 2\langle\Delta\ell_0(t)\Delta\ell_{\rm loss}(t)\rangle = {\rm o}(t)
\end{split}
\mylabel{eq7d}
\end{equation}
The first of the relations (\ref{eq7d}) means that $\Delta\ell_{\rm loss}(\ell',t;\ell)$ is non-zero only if
$\ell',\ell\sim1$ in the limit $t\to0$. The second relation means either
\begin{equation}
 \langle[\Delta\ell_{\rm loss}(t)]^2\rangle ,\langle\Delta\ell_0(t)\Delta\ell_{\rm loss}(t)\rangle = {\rm o}(t)
\mylabel{eq7e}
\end{equation}
or
\begin{equation}
 \Delta\ell_{\rm loss}(t) = - 2\Delta\ell_0(t) + {\rm o}(\sqrt{t})
\mylabel{eq7f}
\end{equation}
The relation (\ref{eq7f}) does not make sense physically, so in what follows we accept option (\ref{eq7e}) and show
that it is consistent with the later calculations.

Next we lift the restriction $t\ll 2/\sigma^2$. It can be shown that the conditional moments (with the condition that
the system was in the state $\ell$ at the beginning of the observation interval) can be expressed as follows:
\begin{equation}
\begin{split}
 & m_{\rm loss}^{(k)}(t;\ell)
 = k!r_{\rm loss}^k \prod_{i=1}^{k} \int\limits_{0}^{t_{i+1}}\!{\rm d}t_i
    \prod_{j=1}^{k-1} w(1,t_{j+1} - t_j;1)
\\
&\qquad\qquad\quad\times
    w(1,t_1;\ell)\ ,\quad t_{k+1}\equiv t
\end{split}
\mylabel{eq11}
\end{equation}
where $w(\ell',t;\ell)$ is determined by (\ref{eq1a}) and
\begin{equation}
\begin{split}
 r_{\rm loss}
  &\equiv
    \lim_{t\to 0}\frac{1}{t}\int\!{\rm d}\ell'\int\!{\rm d}\ell\ \Delta\ell_{\rm loss}(\ell',t;\ell)
\\
  &=
    \lim_{t\to 0}\frac{1}{t}\int\limits_{-\infty}^1\!{\rm d}\ell'{\rm d}\ell\ (\ell'-\ell-at)w_0(\ell',t;\ell)
  = \frac{\sigma^2}{2}
\end{split}
\mylabel{eq12}
\end{equation}
For unconditional moments in the stationary regime we have
\begin{equation}
\begin{split}
 m_{\rm loss}^{(k)}(t)
  &\equiv \int\limits_0^1\!{\rm d}\ell\ m_{\rm loss}^{(k)}(t;\ell) p(\ell)
\\
  &= k! \prod_{i=1}^{k} \int\limits_{0}^{\tau_{i+1}}\!{\rm d}\tau_i
    \prod_{j=1}^{k-1} w(1,t_{j+1} - t_j;1)\cdot p(1)
\end{split}
\mylabel{eq13}
\end{equation}
where $\tau$ is defined in (\ref{eq1b}) and $p(\ell)$ is the stationary solution of (\ref{FP}):
\begin{equation}
 p(\ell) = \frac{2v{\rm e}^{2v\ell}}{{\rm e}^{2v}-1}
\mylabel{eq14}
\end{equation}

To calculate $m_{\rm loss}^{(k)}(t)$ we consider its Laplace transform:
\begin{equation}
\begin{split}
 M_{\rm loss}^{(k)}(\epsilon)
  &\equiv {\cal L}_\tau m_{\rm loss}^{(k)}(t)
   = \int\limits_0^\infty{\rm d}\tau\ {\rm e}^{-\epsilon\tau} m_{\rm loss}^{(k)}(t)
\\
 &= k! p(1) \left[W(1,\epsilon;1)\right]^{k-1} \frac{1}{\epsilon^2}
\end{split}
\mylabel{eq16}
\end{equation}
where $W(1,\epsilon;1)$ is is defined by (\ref{eq2}).
From (\ref{eq16}) we obtain
\begin{equation}
 m_{\rm loss}^{(1)}(t) = p(1) \tau = p(1) \frac{\sigma^2 t}{2}
\mylabel{eq19}
\end{equation}
For the moments (\ref{eq16}) with $k>1$ we can identify the following regimes:
\begin{equation}
 M_{\rm loss}^{(k)}(\epsilon)
  =
  \begin{cases}
   k! p(1) \epsilon^{-(k+3)/2} & \epsilon\gg1 \\
   k! p^k(1) \epsilon^{-(k+1)}  & \epsilon\ll 1
  \end{cases}
\mylabel{eq20}
\end{equation}
Correspondingly, for the moments in $t$-representation we have
\begin{equation}
 m_{\rm loss}^{(k)}(t)
  =
  \begin{cases}
   k! p(1) \displaystyle\frac{\tau^{(k+1)/2}}{\Gamma[(k+3)/2]} & \tau\ll1 \\
   p^k(1) \tau^k & \tau\gg 1
  \end{cases}
\mylabel{eq21}
\end{equation}

Now we turn our attention to the calculation of the PDF $p_{\rm loss}(x;t)$ of the amount of the lost traffic, $x$,
during time $t$. To calculate it we consider its characteristic function in the $\epsilon$-representation:
\begin{equation}
\begin{split}
&
 \tilde P_{{\rm loss}}(s;\epsilon)
  \equiv {\cal L}_x P_{\rm loss}(x;\epsilon)\ ,\
\\
&
 P_{{\rm loss}}(x;\epsilon)
  \equiv {\cal L}_\tau p_{\rm loss}(x;t)
\end{split}
\mylabel{eq31}
\end{equation}
From (\ref{eq31}) we obtain
\begin{equation}
\begin{split}
 \tilde P_{{\rm loss}}(s;\epsilon)
  &= \sum\limits_{k=0}^\infty\frac{(-s)^k}{k!}\int\limits_0^\infty\!{\rm d}x\ x^k {\cal L}_\tau p_{\rm loss}(x;t)
\\
  &= P_{\rm loss}(\epsilon) + \sum\limits_{k=1}^\infty\frac{(-s)^k}{k!} M^{(k)}_{\rm loss}(\epsilon)
\end{split}
\mylabel{eq32}
\end{equation}
where
\begin{equation}
 P_{\rm loss}(\epsilon) = {\cal L}_\tau p_{\rm loss}(t)\ ,\ \
 p_{\rm loss}(t) = \int\limits_0^\infty\!{\rm d}x\ p_{\rm loss}(x,t)
\mylabel{eq32a}
\end{equation}
with $1-p_{\rm loss}(t)$ being the probability for the system not to drop a single packet over the period of time $t$.
Substituting (\ref{eq16}) into (\ref{eq32}) we have
\begin{equation*}
\begin{split}
&
 \tilde P_{{\rm loss}}(s;\epsilon)
  = P_{\rm loss}(\epsilon) + \frac{p(1)}{\epsilon^2}\sum\limits_{k=1}^\infty (-s)^k [W(1,\epsilon;1)]^{k-1}
\\
&\
  = P_{\rm loss}(\epsilon) + \frac{p(1)}{\epsilon^2W(1,\epsilon;1)}
     \left[ - 1 + \frac{1}{1+sW(1,\epsilon;1)} \right]
\end{split}
\end{equation*}
In order that $P_{{\rm loss}}(s;\epsilon)$ did not have an abnormal behaviour (in particular, it did not contain terms
like $\delta(x)$), we must assume that
\begin{equation}
 P_{\rm loss}(\epsilon) = \frac{p(1)}{\epsilon^2W(1,\epsilon;1)}
\mylabel{eq34}
\end{equation}
Hence,
\begin{equation}
 P_{{\rm loss}}(x;\epsilon) = \frac{p(1)}{\epsilon^2W^2(1,\epsilon;1)}
     \exp\left[ -\frac{x}{W(1,\epsilon;1)} \right]
\mylabel{eq35}
\end{equation}
Integrating this relation over $x$, we recover (\ref{eq34}), which shows that our assumption is indeed correct.

In the regimes of short and long times we have
\begin{equation}
 p_{{\rm loss}}(x;t)
  =
  \begin{cases}
   p(1) {\rm erfc}\left[\displaystyle\frac{x}{\sqrt{4\tau}}\right] & \tau\ll1 \\[4mm]
   \delta\Bigl[ x - \tau p(1) \Bigr]  & \tau\gg 1
  \end{cases}
\mylabel{eq36}
\end{equation}
and
\begin{equation}
 p_{{\rm loss}}(t)
  =
  \begin{cases}
   p(1) \displaystyle\sqrt{\frac{4\tau}{\pi}} & \tau\ll1 \\[4mm]
   1  & \tau\gg 1
  \end{cases}
\mylabel{eq36a}
\end{equation}
The conditional PDF (with the condition that the system dropped at least one packet during the time $t$) can be defined
as follows
\begin{equation}
 w_{\rm loss}(x;t)
  \equiv\frac{p_{\rm loss}(x;t)}{p_{\rm loss}(t)}
  =
  \begin{cases}
   \displaystyle\sqrt{\frac{\pi}{4\tau}} {\rm erfc}\left[\displaystyle\frac{x}{\sqrt{4\tau}}\right] & \tau\ll1 \\[4mm]
   \delta\Bigl[ x - \tau p(1) \Bigr]  & \tau\gg 1
  \end{cases}
\mylabel{eq41}
\end{equation}

Now let us  compare the results of the continous approach with those  of Section II. To make the comparison, we calculate the central moments of losses in a similar way as the unconditional ones in Eq.~(\ref{eq13}). Here we will consider  the variance of the
losses $\sigma^2_{\rm loss}(t)$ only in the limit $\tau\gg1$:
\begin{align}\notag
       \sigma^2_{\rm loss}(t)
  &\approx
  m_{\rm loss}^{(1)}(t)
  \left[ \frac{1}{|v|}{\rm cotanh}|v| - \sinh^{-2}|v| \right]
\\&
  \approx
  \begin{cases}
   \displaystyle\frac{2}{3}m_{\rm loss}^{(1)}(t) & |v|\ll 1 \\[4mm]
   \displaystyle\frac{1}{|v|}m_{\rm loss}^{(1)}(t) & |v|\gg 1
  \end{cases}\label{eq24}
\end{align}
In this long-time limit the ratio of the variance to the square of the average vanishes, so that the distribution of data losses obeys the central limit theorem, as also seen from the second line of Eq.~(\ref{eq41}).
This is essentially in agreement with the result of the compressibility $\chi_{\infty}$ in \cite{Yur06}. Naturally, the present considerations are much more general as we have not imposed any artificial limitations on the random input traffic.

Finally, we calculate the correlator of the fluctuations of losses measured during two time intervals of length $t_1$
and $t_2$ correspondingly and separated by the time $T$:
\begin{equation*}
 {\rm corr}(t_1,t_2,T)
  = \int\limits_0^1\!{\rm d}\ell\ \rho(t_1,t_2,T)
   - m_{\rm loss}^{(1)}(t_1)m_{\rm loss}^{(1)}(t_2)
\end{equation*}
where
\begin{equation*}
\begin{split}
 &\rho(t_1,t_2,T)
\\
&\
 = r_{\rm loss}^2\int\limits_{0}^{t_1}\!{\rm d}t'_1\!\int\limits_{0}^{t_2}\!{\rm d}t'_2\
     w(1,t'_1 + t_2 - t'_2 + T;1) p(1)
\end{split}
\end{equation*}
with $r_{\rm loss}$ defined in (\ref{eq12}).

In the regime $T\gg t_1,t_2$ and $T\gg 2/\sigma^2$ it can be shown that
\begin{equation}
 {\rm corr}(t_1,t_2,T)
  \mathop{\rightarrow}\limits_{T\to\infty}0\ ,
\mylabel{eq48}
\end{equation}
as we would expect. In fact, the correlator goes to zero exponentially if $v\neq0$. In the opposite regime
$2/\sigma^2\gg T\gg t_1,t_2$ we have
\begin{equation}
 {\rm corr}(t_1,t_2,T)
  = m_{\rm loss}^{(1)}(t_1)m_{\rm loss}^{(1)}(t_2)
  \frac{1}{p(1)}\sqrt{\frac{2}{\pi\sigma^2 T}}
\mylabel{eq49}
\end{equation}
which is again in agreement with the results of the discrete-time considerations of Section II, showing the universality of the present approach.
\section{Discussion and Conclusion}

As one would expect intuitively, loss events separated widely in time are uncorrelated as shown by
equation~(\ref{eq48}). By widely separated in time, we mean that the time separation of the two observation intervals
in which losses occur is much longer than the time over which fluctuations of queue length become comparable or much
bigger than the buffer size itself, i.e. $2/\sigma^2$.

However, in the case when the separation time is much smaller than $2/\sigma^2$, the correlations of loss fluctuations
are decaying very, very slowly, as can be seen from equation~(\ref{eq49}). Such time intervals are likely to be
comparable or even smaller than the round trip times for typical TCP connections. TCP is the protocol that controls the
rate at which data is sent across a network, between a particular source and destination. The exact details of the
congestion control operation of TCP can be found in \cite{RFC2581}.

Considerations of losses in a network, rather than in a single buffer, would require knowledge of the distribution and correlations of data traffic through different buffers comprising the nodes. The two input parameters, $a$ and $\sigma^2$ in Eq.~(\ref{moments}) for a single buffer, are determined by the network topology, the routing protocol, and the external input traffic distribution to the network. Of course, a detailed knowledge of all these parameters is never available for a realistic network. We will consider an analytically tractable albeit a simplified model with a homogeneous external traffic (all flows from any source to any destination are considered statistically equivalent). Then the above individual single-buffer input parameters are straightforwardly connected to the number of flows passing through the appropriate buffer. This number, in turn, depends on the topology, the protocol and the external load and is equal to the link-betweenness of the corresponding buffer. Fortunately for our considerations the distribution of these quantities are empirically known through measurements on the Internet \cite{Krukov}. This allowed us to analyze fluctuations and temporal correlations of losses in a realistic model of the Internet \cite{ASS}.

\section*{Acknowledgement}

This work was supported by the EPSRC grant GR/T23725/01.


\begin{thebibliography}{10}

\bibitem{Arbib:01}
M.~A. Arbib, {\it The Handbook of Brain Theory and Neural Networks}, MIT Press, London (2003).

\bibitem{Jeong:00}
H.~Jeong, B.~Tombor, R.~Albert, Z.~N. Oltvai, and A.-L. Barab{\' a}si, {\it
  Nature} {\bf 407}, 651 (2000).

\bibitem{Cohen:90}
J.~E. Cohen, F.~Briand, and C.~M. Newman, {\it Community Food Webs: data and
  theory}, Springer Verlag, Berlin (1990).

\bibitem{Liljeros:01}
F.~Liljeros, C.~R. Edling, L.~A.~N. Amaral, H.~E. Stanley, and Y.~Aberg, {\it
  Nature} {\bf 411}, 907 (2001).

\bibitem{Pastor:01}
R.~Pastor-Satorras and A.~Vespignani, {\it Phys. Rev. Lett.} {\bf 86}, 3200
  (2001).

\bibitem{ErdRen}
P.~Erd\"{o}s and A.~R\'{e}nyi, {\it Publ. Math. I. Hung.} {\bf 5}, 17 (1960).

\bibitem{Barabasi:99}
A.-L. Barab{\' a}si and R.~Albert, {\it Science} {\bf 280}, 98 (1999).

\bibitem{Watts:98}
D.~J. Watts and S.~H. Strogatz, {\it Nature} {\bf 393}, 440 (1998).

\bibitem{Watts:99}
D.~J. Watts, {\it Small worlds: the dynamics of networks between order and
  randomness}, Princeton UP (1999).

\bibitem{Albert:02}
R.~Albert and A.-L. Barab{\' a}si, {\it Rev. Mod. Phys.} {\bf 74}, 47 (2002).

\bibitem{Albert:00}
R.~Albert, H.~Jeong, and A.-L. Barab{\' a}si, {\it Nature} {\bf 406}, 378
  (2000).

\bibitem{Cohen:00}
R.~Cohen, K.~Erez, D.~{ben-Avraham}, and S.~Havlin, {\it Phys. Rev. Lett.} {\bf
  85}, 4626 (2000); {\it ibid}
{\bf
  86}, 3682 (2001).

\bibitem{Braunstein:03}
L.~A. Braunstein, S.~V. Buldyrev, R.~Cohen, S.~Havlin, and H.~E. Stanley, {\it
  Phys. Rev. Lett.} {\bf 91}, 168701 (2003).

\bibitem{Dorogovtsev:00}
S.~N. Dorogovtsev and J.~F.~F. Mendes, {\it Europhys. Lett.} {\bf 50}, 1
  (2000).

\bibitem{Ashton:05}
D.~J. Ashton, T.~C. Jarrett, and N.~F. Johnson, {\it Phys. Rev. Lett.} {\bf
  94}, 058701 (2005).

\bibitem{Moreno:03}
Y.~Moreno, R.~Pastor-Satorras, A.~Vázquez, and A.~Vespignani, {\it Europhys.
  Lett.} {\bf 62}, 292 (2003).

\bibitem{Ohira:98}
T.~Ohira and R.~Sawatari, {\it Phys. Rev. {\rm E}} {\bf 58}, 193 (1998); S.~G\`{a}bor and I.~Csabai, {\it Physica A}
{\bf 307}, 516 (2002).

\bibitem{IVL:93}
I.~V. Lerner, {\it Nucl. Phys.} A {\bf 560}, 274 (1993).

\bibitem{Cohen:69}
J.~W. Cohen, {\it Single Server Queue}, North-Holland, Amsterdam (1969).

\bibitem{Schwartz:87}
M.~Schwartz, {\it Telecommunication Networks, Protocols, Modeling and
  Analysis}, Addison-Wesley (1987).

\bibitem{Cates:98}
O.~J. O\char39{}Loan, M.~R. Evans, and M.~E. Cates, {\it Phys. Rev. E} {\bf 58}, 1404 (1998); T.~Nagatani, {\it ibid}
{\bf 58}, 4271 (1998).

\bibitem{Menezes}
M.~A. de~Menezes and A.-L. Barab{\' a}si, {\it Phys. Rev. Lett.} {\bf 92},
  028701 (2004);
J.~Duch and A.~Arenas, {\it ibid} {\bf 96}, 218702 (2006).

\bibitem{Yur06} I.V. Yurkevich, I.V. Lerner, A.S. Stepanenko and C.C. Constantinou, {\it Phys. Rev. E} { \bf 74}, 046120
(2006).

\bibitem{RFC2581} M. Allmanm, V. Paxson and W. Stevens, Internet RFC 2581, IETF
(1999).

\bibitem{Krukov}
X. Dimitropoulos, D. Krioukov, and G. Riley. \textit{Revisiting internet aslevel
topology discovery}. In ``Passive and Active Measurement Workshop''
(PAM), Boston, MA,  (2005).


\bibitem{ASS} A.S. Stepanenko, C.C. Constantinou, I.V. Yurkevich, and I.V. Lerner, {\it in prepartion}.
(2008).
\end{thebibliography}
\end{document}